\documentclass{aa}
\usepackage{graphicx}
\input psfig.sty
\begin{document}
   \title{Deep \hbox{\ion{H}{i}}~ observations of the compact high-velocity cloud
   \object{HVC125+41--207}}

   \subtitle{}

   \author{C. Br\"uns\inst{1,2}
         \and J. Kerp\inst{1}
         \and A. Pagels\inst{1}
          }

   \institute{Radioastronomisches Institut der Universit\"at Bonn, 
              Auf dem H\"ugel 71, D-53121 Bonn, Germany \\
         \and
   The Australia Telescope National Facility, CSIRO, 
       PO Box 76, Epping NSW 1710, Australia\\
             }
       \offprints{C. Br\"uns, Bonn address,
       \email{cbruens@astro.uni-bonn.de}}
   \date{Received 14 December 2000; accepted 7 March 2001}

   \abstract{We present deep \hbox{\ion{H}{i}}~ observations of the compact high-velocity 
   cloud \object{HVC125+41--207} using the 100-m Effelsberg telescope. 
   Our goal was in particular to study the warm neutral medium (WNM) in detail.
   The Effelsberg data reveals a two phase core/halo structure -- one component 
   with a velocity width of FWHM $\approx$ 5 $\rm km~s^{-1}$ (Westerbork data 
   show FWHM $\approx$ 2 $\rm km~s^{-1}$, Braun \& Burton \cite{bb2000}) and 
   one with FWHM $\approx$ 18 $\rm km~s^{-1}$. 
   The column density distribution of the warmer component is highly asymmetric 
   and shows a head-tail structure. We performed a Gaussian decomposition of 
   the cloud and found that 52\% of the \hbox{\ion{H}{i}}~ mass of the cloud is in the WNM. 
   24\% of the WNM is located in the tail. 
   The overall structure and the systematic
   variation of the observational parameters radial velocity, velocity
   dispersion and column density indicate that this cloud is currently
   interacting with the ambient medium. \\
   The Westerbork \hbox{\ion{H}{i}}~ data of this HVC (Braun \& Burton 
   \cite{bb2000}) reveals an interesting dense condensation. Assuming that this 
   condensation is virialized and in pressure equilibrium with the ambient medium, we derive
   a distance of 130 kpc for \object{HVC125+41--207}. 
   Following these considerations, it is possible to constrain the
   parameters $n_{\rm IGM}~<~$7.8$\cdot$10$^{-6}$ cm$^{-3}$ and 
   $T_{\rm IGM}$ $>$ 1.1$\cdot$10$^{5}$K of the intergalactic medium of the 
   Local Group.
   }

   \maketitle
%
%________________________________________________________________

\section{Introduction}

High-velocity clouds (HVCs), first discovered by Muller et al. 
(\cite{muller}), are defined as neutral atomic hydrogen clouds with  
radial velocities (relative to the local--standard--of--rest frame, LSR) that
can not be explained by simple galactic rotation models. Up to now there is no 
general consensus on the origin and the basic physical parameters of HVCs. The 
most critical issue of HVC research is the distance uncertainty. 
Danly et al. (\cite{danly}) determined an upper limit of the distance to
the Galactic Plane of $z$ $\leq$  3.5 kpc for HVC complex M. 
The most important step forward is the bracket of the distance from the Galactic 
Plane of 2.5 $\leq z \leq$ 7 kpc for HVC complex A (van Woerden 
et al. \cite{van Woerden}).
These results clearly indicate that the HVC complexes M and A are located in the
gaseous halo of the Milky Way.

Oort (\cite{oort}) proposed an extragalactic origin for the
HVCs. He argued, that the formation of galaxies is still an ongoing process and 
HVCs represent primordial clouds that are currently accreted by the
Milky Way.
Blitz et al. (\cite{blitz}) revived the hypothesis that some HVCs are 
primordial gas left over from the formation of the Local Group galaxies.
Braun \& Burton (\cite{bb99}) identified 65 compact and isolated HVCs
and argued that this ensemble represents a homogeneous subsample of HVCs at
extragalactic distances.
Observational evidence for extragalactic HVCs may also be found by the detection of
the highly ionized high-velocity gas by Sembach et al. (\cite{sembach}),
 because of its very low pressure of about 2 K ${\rm cm^{-3}}$. 

Meyerdierks (\cite{meyerdierks}) detected an HVC that appears like a cometary 
shaped cloud with a central core and an asymmetric envelope of warm neutral atomic 
hydrogen. He
interpreted this head-tail structure as the result of an interaction between 
the HVC and normal galactic gas at lower velocities.
Towards HVC-complex C, Pietz et al. (\cite{pietz96}) discovered the so-called 
\hbox{\ion{H}{i}}~ ``velocity bridges'' which seem to connect some HVCs with the normal 
rotating interstellar medium.
The most straight forward interpretation for the existence of such
structures is to assume that a fraction of the HVC gas was 
stripped from the main condensation.
Br\"uns et al. (\cite{hvcpaper}) extended the investigations of 
Meyerdierks (\cite{meyerdierks}) and Pietz et al. (\cite{pietz96})
over the entire sky covered by the new Leiden/Dwingeloo \hbox{\ion{H}{i}}~ 21-cm line
survey (Hartmann \& Burton \cite{hartmann97}) and found head-tail structures in
all HVC-complexes except for the very faint HVC-complex L. Their analysis
revealed that the absolute value of the radial velocity of the tail is always 
lower than the value for the head of the
HVC ($|v_{\rm LSR,tail}| < |v_{\rm LSR,head}|$). In addition, it was shown that the fraction of HVCs showing a head-tail
structure increases proportional to the peak column density and increasing 
radial velocity $|v_{\rm GSR}|$.

In this {\em letter}, we present deep integrated Effelsberg \hbox{\ion{H}{i}}~ 21-cm line data 
of the compact high-velocity cloud \object{HVC125+41--207} (discovered by Hulsbosch \&
Wakker 1988), which was placed at
extragalactic distances by Braun \& Burton (\cite{bb2000}). In Sect. 2,
we give a brief summary of the observational parameters. The observations show
that \object{HVC125+41--207} is a head-tail HVC. The results -- including a Gaussian 
decomposition of the spectra -- are presented in Sect. 3. 
We summarize and discuss the results of the Gaussian decomposition and interpret
the head-tail structure as an interaction with an ambient medium in Sect. 4. In
addition we derive the distance to \object{HVC125+41--207} and
constrain the parameters $n$ and $T$ of the intergalactic medium.  

%__________________________________________________________________

\section{Observations}
The observations were carried out between June and October 2000 at the
100-m Effelsberg telescope using the 21-cm receiver (T$_{\rm sys} \approx$ 24 K). 
The HPBW at 21-cm is 9'. The 1024 channel autocorrelator was splitted into two 
banks of 512 channels for each polarization.
The bandwidth of 1.5 MHz offers a velocity resolution of 0.65 $\rm km~s^{-1}$~. 
The HVC was mapped on a regular grid in equatorial coordinates with 
grid-spacings of 4.5 arcmin and an integration time of 4 minutes per 
position. The standard calibration source S7 was used for the flux-calibration. 
We subtracted a third order polynomial for the baseline correction. After 
averaging the two polarizations we reach an rms-noise of 0.09 K at full velocity 
resolution of 0.65 $\rm km~s^{-1}$~. \\
Radio frequency interference was reduced significantly during the last two years
at the Effelsberg telescope. There is no detectable interference within the used
bandwidth.

%The data from the Westerbork interferometer, shown in Fig. \ref{hvccolumn}, were
%observed by Braun \& Burton (\cite{bb2000}). The angular and velocity resolution 
%are 28 arcsec and 2 $\rm km~s^{-1}$~, respectively. More details can be found in 
%Braun \& Burton (\cite{bb2000}).

%__________________________________________________________________

\section{Results}

\begin{figure*}[]
  \centerline{
\psfig{figure=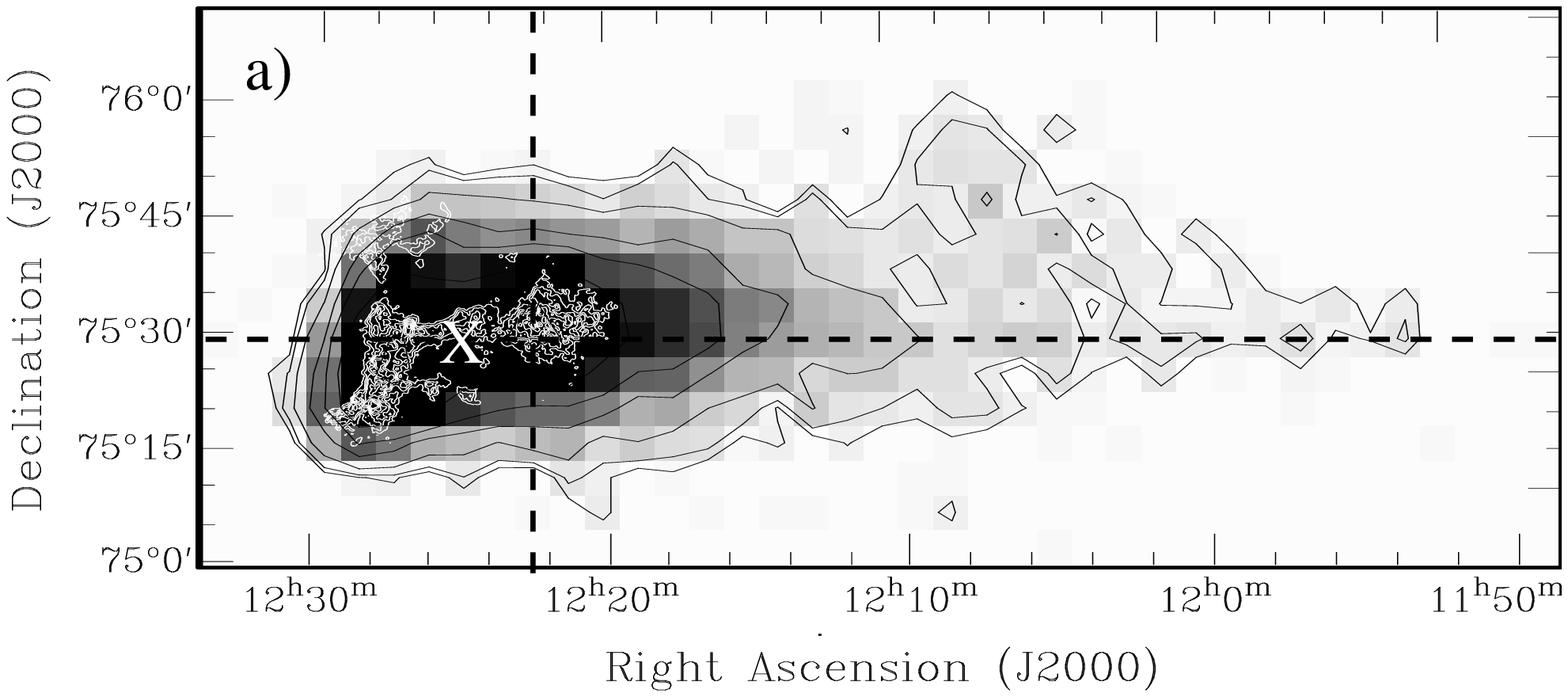,height=7.7cm,angle=0,bbllx=0pt,bblly=15pt,bburx=590pt,bbury=270pt}}
  \centerline{
  \psfig{figure=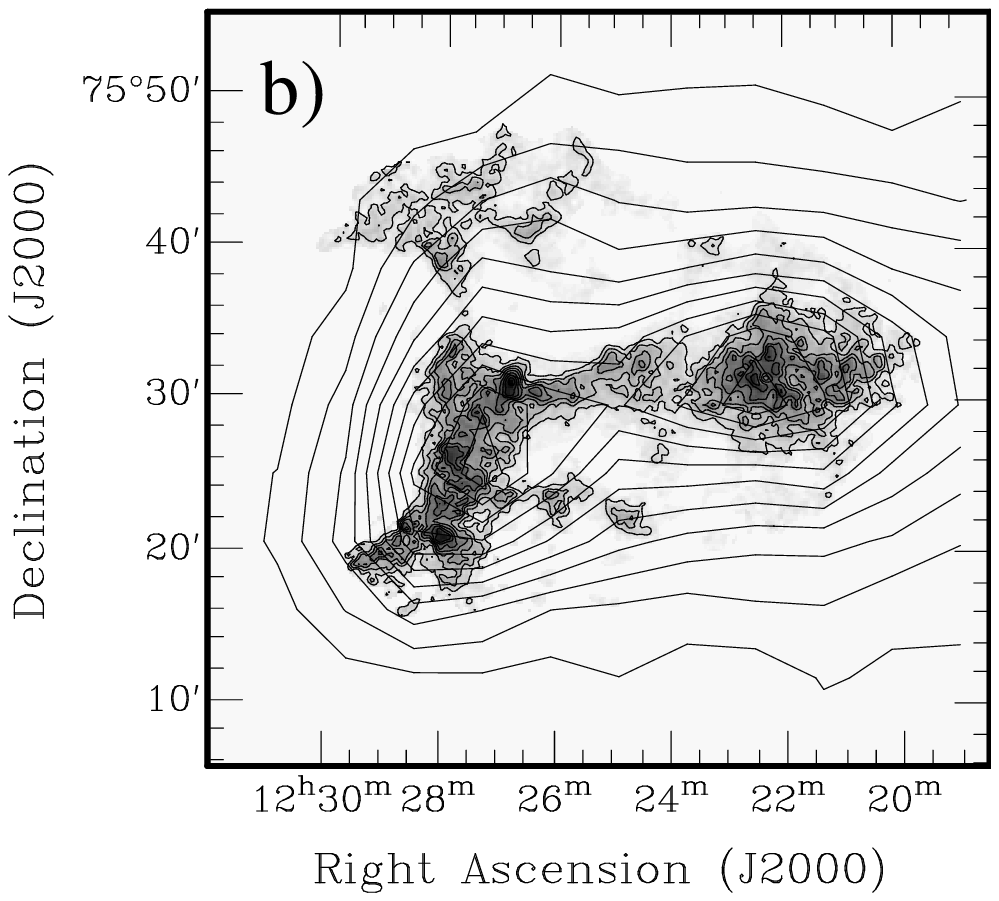,height=7.0cm,angle=0}
  \hfill
  \psfig{figure=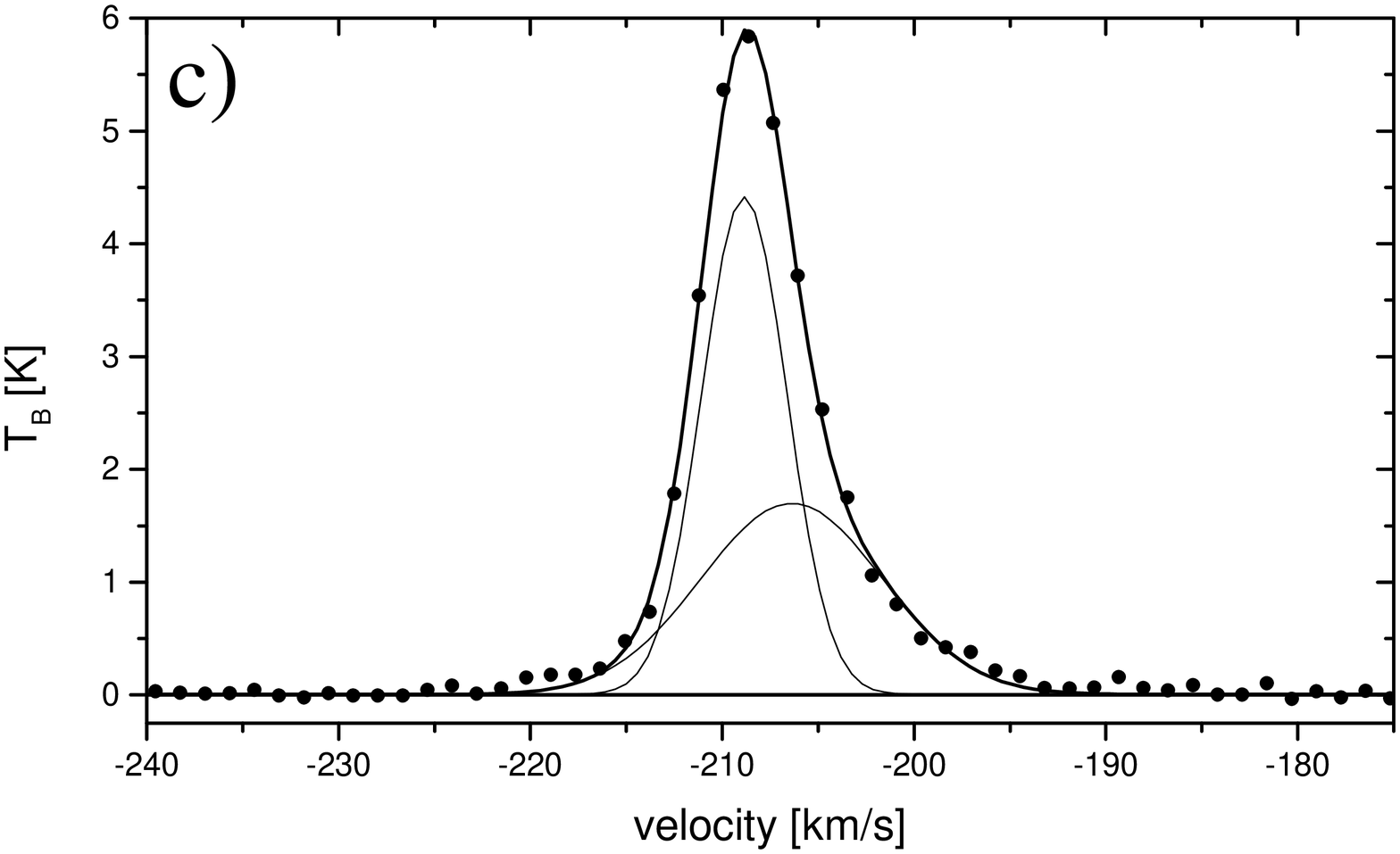,height=7.8cm,width=10.0cm,angle=0,bbllx=0pt,bblly=270pt,bburx=750pt,bbury=750pt}}
 \caption[]{{\bf a} Column density distribution of \object{HVC125+41--207}. The black 
 contour levels represent 3, 5, 10, 20, 30, 50, 100$\cdot10^{18} {\rm cm}^{-2}$
 as observed with the Effelsberg telescope. The white contour levels represent
 the Westerbork data from Braun \& Burton (\cite{bb2000}) starting from 
 50$\cdot10^{18} {\rm cm}^{-2}$ in steps of 50$\cdot10^{18} {\rm cm}^{-2}$.  
 The distribution of the Effelsberg \hbox{\ion{H}{i}}~ data show a head-tail structure. 
 The dashed lines represent the slices through the cloud (see Fig. \ref{gauss}).
 The ``X'' indicates the position of the spectrum shown in Fig. 1c. 
 {\bf b} column density distribution of the Westerbork data with same contour
 levels as in Fig. 1a. The Effelsberg data are represented by contour levels
 starting with 5 $\cdot10^{18} {\rm cm}^{-2}$ increasing in steps of 
 10 $\cdot10^{18} {\rm cm}^{-2}$.
 {\bf c} This figure shows a typical \hbox{\ion{H}{i}}~ spectrum containing a warm and a cold
 component (represented by the two Gaussians). The points represent the data, 
 the thick line is the fit to the spectrum. Note that the warm gas-phase is at 
 a lower radial velocity than the cold component.}
 \label{hvccolumn}
\end{figure*}

Fig. \ref{hvccolumn}a shows the column density distribution of the compact
high-velocity cloud \object{HVC125+41--207}. Black contour lines represent the Effelsberg
observations. The lowest contour line indicates a column density of 
3$\cdot$10$^{18}$cm$^{-2}$. This corresponds to 3$\sigma$ significance level. 
The peak column density in this map is 1.2$\cdot$10$^{20}$cm$^{-2}$. 
The distribution of the low column density material is highly asymmetric. While
there is a sharp column density gradient on the eastern side, the column density
on the western side drops off very slowly. This kind of morphology is called 
head-tail structure (Meyerdierks \cite{meyerdierks}, Br\"uns et al.
\cite{hvcpaper}).

\subsection{Gaussian decomposition}

Fig. \ref{hvccolumn}c shows a typical \hbox{\ion{H}{i}}~ spectrum containing one component with 
a low and one with a high velocity dispersion.
The two component structure is visible in all spectra with column densities
N$_{\hbox{\ion{H}{i}}~} \ge$ 2$\cdot$10$^{19}$ cm$^{-2}$, consistent with 
HVC-models (Wolfire et al. \cite{wolfirea}) and previous observations.

We performed a Gaussian decomposition of the Effelsberg dataset 
to study the two gas-phases in detail. Each spectrum was inspected individually 
if a single Gaussian fits the shape of the line sufficiently. 
If not, two Gaussians were fitted to the spectrum. In the very end of the tail 
($\alpha_{2000} <$ 12$^{\rm h}$05$^{\rm m}$) neighbouring spectra were 
averaged and hanning smoothed to gain a higher signal-to-noise ratio.

\subsection{The column density distribution}

The Gaussian decomposition gives the opportunity to separate the large and the
low velocity dispersion components. 
The low velocity dispersion component is concentrated in two main condensations, 
clump I centred on ($\alpha_{2000}$ = 12$^{\rm h}$27$^{\rm m}$20$^{\rm s}$, 
$\delta_{2000}$ = 75\degr25'30") 
and clump II centred on ($\alpha_{2000}$ = 12$^{\rm h}$22$^{\rm m}$30$^{\rm s}$, 
$\delta_{2000}$ = 75\degr31'30").
The two clumps have similar peak column densities 
(N$_{\hbox{\ion{H}{i}}~} \approx$ 8$\cdot$10$^{19}$ cm$^{-2}$). 

The column density distribution of the component with the large velocity
dispersion also shows two maxima, but shifted in respect to the cold gas-phase 
(see Fig. \ref{gauss}a). The column density distribution is highly asymmetric, 
in the tail it decreases exponentially (e$^{-\alpha/\tau}$ 
with $\tau$ = 0\fdg515 $\pm$ 0\fdg014). 
The last four data points deviate significantly from an exponential decay.

The slice perpendicular to the cloud (Fig. \ref{gauss}b) shows that the warm 
component is smoother distributed and more extended (FWHM = 18') than the 
cold component (FWHM = 11'). 
The Westerbork data (Fig. \ref{hvccolumn}b) show, that the extent of the cold
component (FWHM = 6') is much smaller at high angular resolution. 

The total \hbox{\ion{H}{i}}~ mass can by calculated by summing up all Gaussian components over 
the extent of the cloud and assuming a distance. The \hbox{\ion{H}{i}}~ mass scales 
quadratically with distance. 
The total \hbox{\ion{H}{i}}~ mass is $M_{\hbox{\ion{H}{i}}~}$ = 2.4$\cdot$10$^6$ (d/200kpc)$^2$
M$_{\sun}$. 
The component with the large velocity dispersion accounts for 52\% of the 
total \hbox{\ion{H}{i}}~ mass. 24\% of this material (12.5\% of the total \hbox{\ion{H}{i}}~ mass) is 
located in the tail $\alpha_{2000} <$ 12$^{\rm h}$12$^{\rm m}$. 

\subsection{The distribution of temperature and turbulence}

\begin{figure}[]
  \centerline{
  \hfill
\psfig{figure=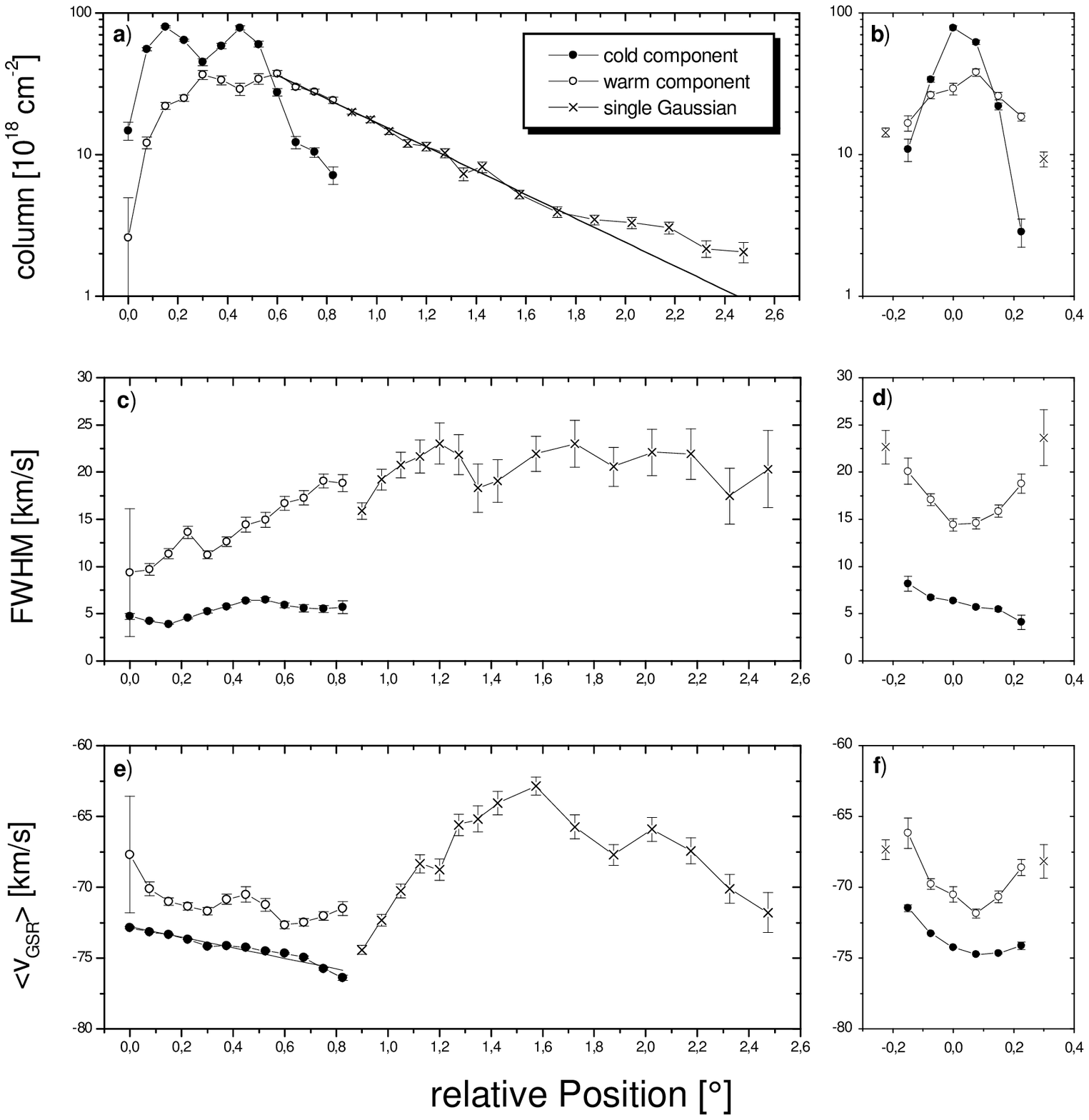,height=9.3cm,angle=0,bbllx=30pt,bblly=90pt,bburx=540pt,bbury=595pt}
  }
 \caption[]{Results of the Gaussian decomposition of the \hbox{\ion{H}{i}}~ spectra.
 The cold gas-phase is represented by filled circles, the warm gas-phase by open
 circles. Results from a single Gaussian fit are represented by crosses. The
 diagrams {\bf a},{\bf c} and {\bf e} show the results for a slice along the 
 tail, the diagrams {\bf b},{\bf d} and {\bf f} represent the 
 slice perpendicular to the orientation of the tail. The line in 
 diagram {\bf a} is an exponential fit to the data.
 The line in diagram {\bf e} is a fit to the velocity field.}
 \label{gauss}
\end{figure}

Obviously, there are two gas-phases, one centered at FWHM = 5 $\rm km~s^{-1}$~ and one centered at 
FWHM = 18 $\rm km~s^{-1}$~. This corresponds to Doppler temperatures of $T_{\rm D}$ = 540 K and 
$T_{\rm D}$ = 7000 K. These are only upper limits to the kinetic temperatures, as
unresolved small scale structure and turbulence also contribute to the Doppler
temperature. The Westerbork observations from Braun \& Burton (\cite{bb2000})
show, that the compact cores have velocity dispersions of the order 
FWHM = 2 $\rm km~s^{-1}$~ or $T_{\rm D}$ = 85 K at an angular resolution of 28". 
A detailed analysis of the small-scale structure is not possible with the
Effelsberg data. Nevertheless, the comparison of the Westerbork and Effelsberg
data (Fig. 1b) shows that the general distribution of the cold component 
can be studied, especially with respect to velocity gradients (see Sect. 3.4).

The warm component is not affected by unresolved small-scale structure. 
However, the data can not distinguish between an increase in turbulence or 
temperature.
The velocity dispersion of the warm gas-phase is increasing from 
FWHM $\approx$~10 $\rm km~s^{-1}$~ to FWHM $\approx$~20 $\rm km~s^{-1}$~ over 
the extend of the cloud and remains approximately constant in the tail 
(see Fig. \ref{gauss}c). This corresponds to an increasing Doppler temperature 
from $T_{\rm D}$ = 2000 K to $T_{\rm D}$ = 10000 K in the tail. 
The slice perpendicular to the cloud shows, that the velocity dispersion 
increases significantly towards the border of the cloud (see Fig. \ref{gauss}d).
For this particular slice, the Doppler temperature increases from 
$T_{\rm D} \approx$ 5000 K in the centre to $T_{\rm D} \approx$ 11000 K at the
border of the cloud.  
 
\subsection{The velocity field}

The interpretation of the velocity gradients in the local-standard-of-rest 
frame is difficult, because the angle between the solar velocity vector and 
the line of sight varies over the extend of the cloud and therefore introduces 
a velocity gradient that is not related to the HVC itself. The effect is of 
the order 2 $\rm km~s^{-1}$~ per degree and can be removed by converting the velocity 
into the galactic-standard-of-rest frame. 

Fig. \ref{gauss}e shows the parameter mean velocity (v$_{\rm GSR}$) along
a slice in direction of the tail. There is a regular velocity gradient in the 
cold phase. This gradient can be easily explained by the same type of projection 
effect that urged us to convert the radial velocity into the 
galactic-standard-of-rest-frame: the angle between the line of sight and the 
HVC velocity vector is changing over the extend of the HVC. 
This effect can be used to derive the three-dimensional velocity of
the HVC. The gradient should be of the form $v_{\rm GSR}$ = $v_{\rm total}$
cos($\alpha$ + $\alpha_0$). The result of a least-squares fit is also shown in 
Fig. \ref{gauss}e. The total space-velocity is $v_{\rm total}$ = 225$\pm$18 $\rm km~s^{-1}$~ and
the angle between the line of sight and the HVC velocity vector is $\alpha_0$ =
108$\fdg$4$\pm$1$\fdg$5, i.e. the direction of proper motion is almost 
perpendicular to the line of sight in the direction of the sharp velocity 
gradient. This result indicates, that the tail is trailing the HVC. 

The spectrum in Fig. \ref{hvccolumn}c shows, that the warm component has a lower 
radial velocity than the cold component. It is remarkable, that this is true 
for {\em all} spectra where a two component fit was feasible. This result 
indicates that the two gas-phases are separating from each other. The difference
between the radial velocities of the warm and the cold component is minimal
where the column density of the warm gas-phase has its maxima. 

The absolute value $|v_{\rm LSR}|$ of the radial velocity is steeply 
decreasing in the first part of the tail and increasing towards the end 
(Fig. \ref{gauss}e). The slice perpendicular to the cloud shows, that 
$|v_{\rm LSR}|$ decreases significantly towards the borders 
of the cloud (see Fig. \ref{gauss}f). 

%__________________________________________________________________

\section{Summary and Discussion}

Our Effelsberg \hbox{\ion{H}{i}}~ observations of \object{HVC125+41--207} show, that the compact HVC 
has a cometary appearance. The \hbox{\ion{H}{i}}~ gas gets warmer and/or more turbulent to the
\hbox{\ion{H}{i}}~ boundaries of the cloud as well as towards the tail. The radial velocity 
$|v_{\rm LSR}|$ of
the \hbox{\ion{H}{i}}~ gas is decreasing towards the border of the cloud and in the tail. 
It is remarkable that the warm gas-phase {\em always} shows lower $|v_{\rm LSR}|$ 
than the cold phase.  

These results demonstrate that \object{HVC125+41--207} is definitively not a relaxed 
cloud in equilibrium. 
The observational results suggest that the warm gas-phase is stripped
off the main body of the HVC.  This kind of behaviour is for instance expected 
from a ram-pressure interaction: the diffuse warm gas has a much larger 
cross-section for the ram-pressure interaction than the small, dense clumps of 
the cold medium. 

Santillan et al. (\cite{santillan}) for instance performed 
MHD simulations to explain head-tail HVCs at $z$-height of a few kpc. They were
successful in explaining head-tail structures. The tail consists of material
that was stripped off the main body (i.e. the head) of the cloud. Their models
showed that the evolution of the shocked layer generates a tail that oscillates,
creating vorticity and turbulent flows along its trajectory. This explains the
velocity structure in the tail of \object{HVC125+41--207} with increasing $|v_{\rm LSR}|$
to the end of the tail. 

It is quite easy to produce a head-tail structure in the lower Galactic
halo. Nevertheless, it might be possible to explain the existence of head-tail 
structures even in the intergalactic space of the local group. Br\"uns et al.
(\cite{hvcpaper}, \cite{mexpaper}) showed that there are head-tail HVCs in the
Magellanic Stream and the Leading Arm. 
New magnetohydrodynamical simulations considering the critical velocity effect  
(Konz, priv. comm.) explain head-tail structures if the ambient medium has
a plasma density of the order $n \approx$ 10$^{-5}$ cm$^{-3}$ and a magnetic 
field $|B|$ of the order of a few $\mu$G. 

The Westerbork data (Fig. \ref{hvccolumn}b) show one remarkable condensation 
($\alpha_{2000}$ = 12$^{\rm h}$26$^{\rm m}$50$^{\rm s}$, 
$\delta_{2000}$ = 75\degr31'26")
with a brightness temperature $T_{\rm B}$ = 75 K and a velocity width of 
FWHM = 2 $\rm km~s^{-1}$~, i.e. a Doppler temperature of $T_{\rm D}$ = 85 K. This clump 
has therefore a kinetic temperature $T_{\rm kin} \approx$ 80 K.
The clump has an apparent column density of N$_{\hbox{\ion{H}{i}}~} =$ 4.5$\cdot$10$^{20}$ 
cm$^{-2}$, but due to optical depth effects a real column density of the order 
N$_{\hbox{\ion{H}{i}}~} \approx$ 1$\cdot$10$^{21}$ cm$^{-2}$.
We derive a distance to \object{HVC125+41--207} assuming that the clump is
virialized and in pressure equilibrium with the ambient medium. 
If we consider an isothermal sphere of uniform density without magnetic fields,
we can use the equation (Spitzer \cite{spitzer})

\begin{eqnarray}
4\pi R^3 P_0 = \frac{3MkT}{\mu}-\frac{3GM^2}{5R} 
\label{spitzer}
\end{eqnarray}

\noindent $R$ is the radius, $M$ the mass and $T$ the temperature of 
the sphere. $P_0$ is the pressure of the ambient medium and $\mu$ 
the mean mass per particle within the sphere.
The mass $M$ is given by $M$ = 4/3 $\pi$ $R^3$ $\rho$, with 
$\rho$ = $\mu$ N$_{\rm \hbox{\ion{H}{i}}~}$ (2$R$)$^{-1}$, where the column density of the 
clump is N$_{\rm \hbox{\ion{H}{i}}~}$ = 1$\cdot$10$^{21}$ cm$^{-2}$. The radius $R$ = 0.5 
$\theta$ $d$ is calculated by the observed angular diameter of the clump 
$\theta$ = 90" and the distance $d$. 
The external pressure is $P_0$ = k $T_{w}$ N$_{\rm \hbox{\ion{H}{i}}~,w}$ 
 $d^{-1}$ $\theta^{-1}_{\rm w}$, where the column density is 
N$_{\rm \hbox{\ion{H}{i}}~,w}$ = 2$\cdot$10$^{19}$ cm$^{-2}$, the temperature is
T$_{w} <$ 3000 K, and the angular diameter of the ambient medium is
$\theta_{\rm w}$ = 18'. 
Inserting into Eq. 1 and solving for the distance $d$ gives

\begin{eqnarray}
d = \frac{15 k (N_{\rm \hbox{\ion{H}{i}}~} T - \theta~\theta^{-1}_{\rm w} N_{\rm \hbox{\ion{H}{i}}~,w}
T_{\rm w})}
{G \pi \theta N_{\rm \hbox{\ion{H}{i}}~}^2 \mu^2}
\label{dist}
\end{eqnarray}

\noindent The unknown values in Eq. 2 are $T_{\rm w}$ and $\mu$. 
Assuming $\mu$ = 1.25 m$_{\rm H}$, i.e. including the normal amount of Helium,
but without any molecular matter, we get possible 
distances between $d(T_{\rm w}$ = 3000 K) = 126 kpc and $d(T_{\rm w}$ = 
100 K) = 134 kpc. Any molecular gas content would lower the derived distance,
e.g. $\mu$ = 2 m$_{\rm H}$ gives $d$~$\approx$~50~kpc.\\
The density of the ambient WNM at $d$ = 130 kpc is 
$n_{\rm w}$~=~9.8$\cdot$10$^{-3}$~cm$^{-3}$ with pressure $P_0 <$ 29.5 K~cm$^{-3}$.

In the remaining part of this section, we will discuss implications on
the intergalactic medium (IGM). Our Effelsberg data demonstrate, that there {\em must}
be an ambient medium, which is responsible for the head-tail structure.
The observed pressure is $P_0$ = $P_{\rm ram}$ + $P_{\rm IGM}$. 
The ram pressure is $P_{\rm ram}$~=~0.5~$\rho_{\rm IGM}$~$v^2$ = 
$P_0$ -- $P_{\rm IGM} <$ 30 K~cm$^{-3}$. An upper limit to the density of the
ambient medium $\rho_{\rm IGM}$ can be calculated using the space velocity of 
$v$~=~225~$\rm km~s^{-1}$ determined in Sect. 3.4: 
$\rho_{\rm IGM} <$ 1.6$\cdot$10$^{-32}$ kg~cm$^{-3}$ = 240 M$_{\sun}$~kpc$^{-3}$.
With $\rho_{\rm IGM}$~=~1.25~m$_{\rm H}$~$n_{\rm IGM}$ follows that 
$n_{\rm IGM}~<~$7.8$\cdot$10$^{-6}$ cm$^{-3}$.

Assuming that the pressure of 2 K~cm$^{-3}$, derived by Sembach et 
al. (\cite{sembach}) for highly ionized high-velocity clouds, is valid for the
IGM in general, the temperature of the IGM can be calculated. 
Using $P$~k$^{-1}$~=~2.3~$n_{\rm IGM}$~$T_{\rm IGM}$, where 2.3 is the number of particles per hydrogen
atom for an ionized medium including helium, gives $T_{\rm IGM}$~$>$~1.1$\cdot$10$^{5}$K.
$T_{\rm IGM}$ is sufficiently high to keep the IGM ionized (neutral gas would have
been detected in absorption against quasars) and low enough to
be undetectable in X-rays.

However, the derived distance of 130 kpc is based on the assumption of 
equilibrium and no molecular gas content. 
As the HVC is obviously interacting with the ambient medium,
equilibrium conditions are not really justified. The HVC could be located at 
larger distances
having a collapsing clump that might form stars in future -- at a distance
of a Mpc the \hbox{\ion{H}{i}}~ mass is about 40 times larger than the virial mass. 
On the other hand, if the clump contains a significant amount of molecular 
matter, this will localize the HVC in the Galactic Halo.   
  
Our observations demonstrated that both, deep integrated observations of the 
warm neutral medium and high resolution observations of the cold neutral medium
are necessary to obtain further insights to the nature of HVCs. 
Deep \hbox{\ion{H}{i}}~ observations of a number of compact HVCs  
are planed for 2001. These observations will help to determine general
parameters in compact HVCs.

\begin{acknowledgements}
      Part of this work was supported by the German
      \emph{Deut\-sche For\-schungs\-ge\-mein\-schaft, DFG\/} project
      number ME 745/19. We thank our referee W.B. Burton for constructive comments
      and R. Braun \&  W.B. Burton for kindly providing their Westerbork data.
\end{acknowledgements}


\begin{thebibliography}{}

  \bibitem[1999]{blitz} Blitz, L., Spergel, D.N., Teuben, P.J., Hartmann, D.,
 Burton, W. B. 1999, ApJ, 514, 818
   \bibitem[1999]{bb99} Braun, R., Burton, W.B. 1999, A\&A, 341, 437
   \bibitem[2000]{bb2000} Braun, R., Burton, W.B. 2000, A\&A, 354, 853 
   \bibitem[2000a]{hvcpaper} Br\"uns, C., Kerp, J., Kalberla, P.M.W., Mebold, U.
   2000a, A\&A, 357, 120
   \bibitem[2000b]{mexpaper} Br\"uns, C., Kerp, J., Staveley-Smith, L. 2000b, in
   ``Mapping the Hidden Universe: The Universe Behind the Milky Way - The 
   Universe in \hbox{\ion{H}{i}}'', eds. Kraan Korteweg R.C. \& Henning P.A., 
   (ASP Conf. 218), 349
   \bibitem[1993]{danly} Danly, L., Albert, C.E., Kuntz, K.D. 1993, ApJ, 416, L29
   \bibitem[1997]{hartmann97} Hartmann, D., Burton, W. B. 1997,
      ``An atlas of Galactic Neutral Hydrogen Emission'', Cambridge University
      Press
   \bibitem[1988]{hulsbosch} Hulsbosch, A.N.M., Wakker, B.P. 1988, A\&AS, 75, 191
   \bibitem[1991]{meyerdierks} Meyerdierks, H. 1991, A\&A, 251, 269
   \bibitem[1963]{muller} Muller, C.A., Oort, J.H., Raimond, E. 1963, C.R. Acad.
   Sci. Paris, 257, 1661
   \bibitem[1966]{oort} Oort, J.H. 1966, Bull. Astr. Inst. Netherlands, 18, 421
   \bibitem[1996]{pietz96} Pietz, J., Kerp, J., Kalberla, P.M.W. et al. 1996,
      A\&A, 308, L37
   \bibitem[1999]{santillan} Santillan, A., Franco, J., Martos, M., Kim, J. 1999 ,
   ApJ, 515, 657
   \bibitem[1999]{sembach} Sembach, K.R., Savage, B.D., Lu, L., Murphey, E.M. 
   1999, ApJ, 515, 108
   \bibitem[1978]{spitzer} Spitzer, L. 1978, ``Physical Processes in the
   interstellar medium'', Wiley, New York
   \bibitem[1999]{van Woerden} van Woerden, H., Schwarz, U.J., Peltier, R.F.,
   Wakker, B. P., Kalberla, P. M. W. 1999, Nat, 400, 138
   \bibitem[1995]{wolfirea} Wolfire, M.G., McKee, C.F., Hollenbach, D., Tielens,
   A.G.G.M. 1995, ApJ, 453, 673
\end{thebibliography}
\end{document}